\documentclass[aps,twocolumn,showpacs]{revtex4}
\usepackage{graphicx}

\begin{document}
\draft

\title{\bf The Quantum Compass Model on the Square and Simple Cubic
Lattices}

\author{J. Oitmaa and C. J. Hamer}

\affiliation{School of Physics, The University of New South Wales, Sydney
2052, Australia.} 
\begin{abstract}
We use high-temperature series expansions to obtain thermodynamic
properties of the quantum compass model, and to investigate the phase
transition  on the square and simple cubic lattices. On the square
lattice we obtain evidence for a phase transition, consistent with
recent Monte Carlo results. On the simple cubic lattice the same
procedure provides no sign of a transition, and we conjecture that there
is no finite temperature transition in this case.
\end{abstract}
\pacs{PACS Indices:
05.30.-d,75.10.-b,75.10.Jm,75.30.Cr,75.30.Kz \\
\\  \\
(Submitted to  Phys. Rev. B) }
\maketitle

\section{Introduction}

Quantum compass models are spin models in which the nearest-neighbour
exchange coupling has the form $J_{\alpha}S^{\alpha}_iS^{\alpha}_j$
where $\alpha (=x,y,z)$ depends on the direction of the particular link or bond. This
then implies a coupling between the spin space and the physical space of
the lattice. Such models were first introduced, and have been regularly
employed, to describe orbital ordering in various transition metal
compounds (\cite{kugel1982,khomskii2003,brink2004,jackeli2009},
and references therein).

Such models also have applicability to models of $p + ip$ superconducting
arrays \cite{xu2004,nussinov2005} and it has been argued \cite{doucot2005} that such arrays can
provide fault-tolerant qubits for quantum information systems. 

Compass models can be defined in various ways, depending on the
underlying lattice. Exact solutions have been obtained for a
1-dimensional alternating (xx),(zz) model \cite{brzezicki2007} and for a 2-leg ladder
\cite{brzezicki2009}. A remarkable solution has also been found for the honeycomb
lattice with (xx), (yy) and (zz) couplings along the three independent
lattice directions \cite{kitaev2006}.
As far as we are aware, no other exact solutions exist.

In the present paper we consider the spin-1/2 quantum compass model on
the simple cubic lattice, with Hamiltonian
\begin{equation}
H = J_x \sum^{(x)}_{<ij>} \sigma^x_i \sigma^x_j + J_y \sum^{(y)}_{<ik>}
\\
\sigma^y_i \sigma^y_k + J_z \sum^{(z)}_{<il>} \sigma^z_i \sigma^z_l
\label{eq1}
\end{equation}
where the $\sigma^{\alpha}_i$ are Pauli operators, and the sums are,
respectively, over lattice bonds along the x,y,z directions. We will
also consider the square lattice version, where the last term in
(\ref{eq1}) is omitted.

As is well known \cite{nussinov2005}, this model possesses a number of unusual
gauge-like symmetries. As a consequence each energy state has a
macroscopic degeneracy and, consequently, there is no conventionally ordered
magnetic phase at any temperature.
However it has been pointed out that a state of orientational or
`nematic' order is possible, in which the nearest neighbour bonds of
lowest energy lie predominantly along a specific lattice direction. In
the isotropic case of equal interactions $(J_x = J_y = J_z)$ this
represents a spontaneous symmetry breaking. Consequently there may be a
critical point at a temperature $T_c$, above which the system is
disordered, with no preferred direction.

Recent quantum Monte Carlo studies of the isotropic 2-dimensional model
\cite{tanaka2007,wenzel2008,wenzel2010} have found strong evidence for a 
finite temperature critical
point with $kT_c/J = 0.234$ (in our units) with a critical exponent $\nu
\simeq 0.97$, consistent with 2D Ising behaviour. The same authors \cite{wenzel2010}
also
identified a transition in the corresponding classical model, but we do
not consider the classical case in the present work. As far as we are
aware, no investigation of the occurrence of such a critical point, or
its value, has been reported in the 3-dimensional case.

The goal of the present work is to attempt to answer this question. We
employ the method of high-temperature series expansions, which has
proven successful in the past \cite{domb} in obtaining accurate values for
critical temperatures and exponents in a wide variety of classical and
quantum models. The basic idea is to expand the Boltzmann factor
$e^{-\beta H}$ in the partition function in powers of $\beta =1/kT$
\begin{eqnarray}
Z & = & Tr\{e^{-\beta H}\} \nonumber \\
& = & \sum_{r=0}^{\infty} \frac{(-1)^r}{r!} Tr(H^r)\beta^r
\label{eq2}
\end{eqnarray}
The coefficients in this series can be evaluated in a number of (related)
ways.
We use a linked cluster approach \cite{oitmaa2006} in which $lnZ$ is evaluated, as
a series in $\beta$, on a sequence of finite connected clusters of
increasing size, and the cluster contributions are combined appropriately to
give the bulk free energy in the form
\begin{eqnarray}
-\beta F & = & \frac{1}{N} lnZ \nonumber \\
 & = & ln 2 +\sum_{r=2}^{\infty} a_r(J_x,J_y,J_z)\beta^r
\label{eq3}
\end{eqnarray}
with the $a_r$ being multinomial expressions of degree $r$ in the J's.
From this one can immediately obtain a corresponding series for the
specific heat.

However the specific heat has, in most cases, only a weak singularity
and is not well suited to estimation of critical properties. Including
an external field which couples to the order parameter $D$,
\begin{equation}
H = H_0 - hD
\label{eq4}
\end{equation}
where we now write the original Hamiltonian (\ref{eq1}) as $H_0$, allows
calculation of a high temperature series for a generalized
`susceptibility'
\begin{equation}
\chi = \frac{1}{\beta} \lim_{h \rightarrow 0} \frac{\partial^2}{\partial
h^2} (\frac{1}{N} ln Z)
\label{eq5}
\end{equation}
The order parameter $D$ was introduced \cite{wenzel2010} for the 2-dimensional
model as
\begin{equation}
D_{2d} = 
 J_x \sum^{(x)}_{<ij>} \sigma^x_i \sigma^x_j - J_y \sum^{(y)}_{<ik>}
\sigma^y_i \sigma^y_k 
\label{eq6}
\end{equation}
i.e. the difference between the energy of the $x$ and $y$ bonds. We
generalize this for the 3-dimensional model to 
\begin{equation}
D_{3d} = 
 2J_z \sum^{(z)}_{il} \sigma^z_i \sigma^z_l
 - J_x \sum^{(x)}_{<ij>} \sigma^x_i \sigma^x_j - J_y \sum^{(y)}_{<ik>}
\sigma^y_i \sigma^y_k 
\label{eq7}
\end{equation}

Normally the calculation of the susceptibility would be somewhat
involved, since $H_0$ and $D$ do not commute. However, in the present
model , we can simply combine the two terms into a Hamiltonian of the
original form (\ref{eq1}),
with $J_x \rightarrow J_x(1-h)$, $J_y \rightarrow J_y(1-h)$, $J_z
\rightarrow J_z(1+2h)$
and use the expression in (\ref{eq3}) to obtain
\begin{equation}
\beta \chi = \sum_{r=2}^{\infty} c_r(J_x,J_y,J_z) \beta^r
\label{eq8}
\end{equation}
where the $c_r$ are again multinomials of degree $r$ in the J's. The
susceptibility series is expected to show a strong divergence at the
critical point and hence should be more amenable to analysis.

Another quantity which is expected to show a strong divergence is the
fluctuation in the order parameter
\begin{equation}
Q = <D^2> - <D>^2.
\label{eq9}
\end{equation}
For the classical model this quantity is identical to $\chi$, but this
is not the case for the quantum model.

In the following sections we will present the series and our analysis
for the 2-d case (Section \ref{sec2}) and 3-d case (Section \ref{sec3}). Our conclusions
are summarized in Section \ref{sec4}.
 
\section{The Square Lattice}
\label{sec2}

To test the effectiveness of the high-temperature series approach for
the present model, we first investigate the square lattice case, where
previous results exist \cite{tanaka2007,wenzel2008,wenzel2010}.

We use a linked cluster method \cite{oitmaa2006} based on connected clusters
(`graphs'), To obtain a series for $(ln Z)/N$ correct to order
$\beta^{24}$, as we have done, requires the enumeration of clusters with
up to 12 bonds. It is a special feature of this model that each bond
must be used an even number of times to give a nonzero trace. There are
4423 topologically distinct clusters with 12 or fewer bonds, embeddable
on the square lattice. This gives rise to 751663 distinct graphs with 2
bond types ($x$ and $y$). However the vast majority of these do not
contribute, and the final irreducible list of contributing graphs
numbers 60127. We give below the leading terms in the partition function
series
\begin{widetext}
\begin{eqnarray}
 \frac{1}{N} ln Z & = & ln 2 +\frac{1}{2}(x^2 + y^2)\beta^2 
 -\frac{1}{12} (x^4
+ 8x^2y^2 + y^4) \beta^4 \nonumber \\
 & & +\frac{1}{45}((x^6 + y^6) + 30(x^4y^2 +x^2y^4)) \beta^6 
 -\frac{1}{2520}(17(x^8 + y^8) + 1376(x^6y^2 +x^2y^6) +4344x^4y^4 )
 \beta^8 \nonumber \\
 & & +\frac{1}{14175}(31(x^{10} +y^{10})+5570(x^8y^2 +x^2y^8) +40500(x^6y^4 + x^4y^6)
) \beta^{10} \nonumber \\
& & -\frac{1}{935550}(691(x^{12} +y^{12})+ 241800(x^{10}y^2 + x^2y^{10})+3426402(x^8y^4+x^4y^8)
+7679480x^6y^6 ) \beta^{12}
 \cdots
\label{eq10}
\end{eqnarray}
where $x \equiv J_x, y \equiv J_y$.
Note that only even powers of $\beta$ occur. This is a feature of all
series for this model.

From this result we can obtain the susceptibility
\begin{eqnarray}
 \chi/\beta & = & (x^2 + y^2) -\frac{1}{3} (3(x^4+y^4)
- 8x^2y^2 ) \beta^2 
 +\frac{2}{3}((x^6 +y^6)-2(x^4y^2 +x^2y^4))\beta^4 \nonumber \\
 & & -\frac{1}{315}(119(x^8 + y^8)+1376(x^6y^2+x^2y^6) -4344x^4y^4 )
 \beta^6 \nonumber \\
 & & +\frac{1}{14175}(2790(x^{10}+y^{10}) +144820(x^8y^2+x^2y^8) -243000(x^6y^4 +x^4y^6)
) \beta^8 
 \cdots
\label{eq11}
\end{eqnarray}
\end{widetext}

\begin{widetext}
\begin{center}
\begin{table}
\caption{Series coefficients for the isotropic 2d Compass Model}
\begin{tabular}{|cccc|}
\hline
 p &  $\frac{1}{N}ln Z$ & $\chi/(\beta J^2)$ & Q \\
\tableline   
\ 0 \ & \ 0.693147180560D+00 \ & \  0.200000000000D+01 \
& \  0.200000000000D+01 \ 
\\
\ 2 \ & \ 0.100000000000D+01 \ &  \ 0.666666666666D+00\
 &  \ 0.600000000000D+01 \
\\
 \ 4 \ & \ -0.833333333333D+00 \ & \ -0.133333333333D+01\ & \
-0.120000000000D+02 \ 
\\ 
 \ 6 \ &  \ 0.137777777778D+01 \ &  \ 0.429841269841D+01\
 &
 \ 0.307111111111D+02 \ \\
 \ 8 \ & \ -0.282936507937D+01 \ & \ -0.132382287013D+03 \ & \ -0.849650793664D+02
 \ 
\\
 \ 10 \ & \ 0.650455026455D+01 \ &  \ 0.421280743947D+02 \ &
 \ 0.245585890085D+03 \ \\
 \ 12 \ & \ -0.160518048207D+02 \ & \ -0.132382287013D+03 \ & \ -0.730131587978D+03
 \ \\
 \ 14 \ & \ 0.416028785294D+02 \ &  \ 0.418143457749D+03 \ &
 \ 0.221416777392D+04 \ \\
 \ 16 \ & \ -0.111781974764D+03 \ & \ -0.132765859211D+04 \ & \ -0.681447081860D+04
 \ \\
 \ 18 \ &  \ 0.308758184039D+03 \ &  \ 0.423612006027D+04 \ &  \ 0.212139054331D+05
 \ \\
 \ 20 \ & \ -0.871688240896D+03 \ & \ -0.135761546397D+05 \ & \ -0.666457337381D+05
 \ \\
 \ 22 \ &  \ 0.250500394206D+04 \ &  \ 0.436833196403D+05 \ &  \ 0.210941032291D+06
 \ \\
 \ 24 \ & \ -0.730521400959D+04 \ & & \\
\hline
\end{tabular}
\label{tab1}
\end{table}
\end{center}
\end{widetext}
The higher order terms were evaluated numerically.
In Table \ref{tab1} we show the full series for the isotropic case $J_x
= J_y$. The expansion variable is $K=\beta J$.

%\begin{figure}
%\includegraphics[scale=0.4]{fig1.eps}
%\caption{
%Positions in $K$ of poles in Dlog Pad{\' e} approximants to $ 
%\chi/\beta$ for the 2D compass model. Crosses: original series; circles: 
%after Euler transform.
%}
%\label{fig1}
%\end{figure}

We have attempted to analyse these series using standard
Pad{\' e} approximant methods. Our discussion is confined to
the $\chi$ series, as this (together, possibly, with $Q$) is
expected to have a strong singular behaviour at the critical
point. The first point to make about the series in $\beta^2$
is the regular alternation in sign. This reflects the
presence of a dominant singularity on the negative $\beta^2$
axis (i.e. the imaginary $\beta$ axis). In fact there appears
to be a whole string of such imaginary poles in the Dlog
Pad{\' e} approximants. This, in itself, is not so unusual.
Recall that the exact result for the 1D Ising model has
poles at $\beta J = \pm i(n+1/2)\pi$.

However these interfering singularities mask the expected
physical singularity on the real positive $\beta$ axis. One
possible strategy to overcome this is to use an Euler
transformation of the form $y = x/(1+ax), (x = K^2)$,
which has the effect of compressing the positive real axis
and expanding the region $-1/a < x < 0$ of the negative real
axis. The use of such transformations is well known in the
field of critical phenomena, as are the possible pitfalls.

\begin{table}
\caption{
Poles and residues (in brackets) in the variable $K^2$
for [N/D] Dlog Pad{\' e} approximants to the quantity
$\chi/(\beta J^2)$ for the 2D compass model, after an Euler transform
with $a = 2.0$. (Asterisks
denote a complex pair of poles in the physical region)
}
\begin{tabular}{|c|ccccc|}
\hline
 $D\backslash N$ & 3 & 4 & 5 & 6 & 7 \\
\tableline   
 3 & & 0.439(0.191) & * & 0.457(0.288) & * \\ 
 4 & 0.470(0.460) & 0.474(0.533) & 0.472(0.509)  & 0.475(0.567)&  \\
 5 & 0.473(0.529) & 0.473(0.514) & 0.473(0.575) & & \\
 6 & 0.473(0.512) & 0.473(0.524) & & & \\
 7 & 0.474(0.546) & & & & \\
\hline
\end{tabular}
\label{tab2}
\end{table}

To provide the reader with some insight into the analytic
structure of the $\chi$ series we discuss
the location of poles of Dlog Pad{\' e}
approximants to the series for $\beta J^2/\chi$ before and after the Euler transformation (with
a=2.0). 
The original series in $x = \beta^2$ has very consistent poles at $x
\simeq -0.28, \ -0.32, \ -0.46$, with less
consistent poles much further from the origin. The transformed series
shows images of these at $y = -0.65, \ -0.9$ as well as poles on the
positive real axis at $y = 0.47, \ 0.54$. The last of these corresponds
to a large negative value $x \simeq -6.8$, whereas  $y = 0.47$
corresponds to $x = 7.8$, or a physical critical value $kT_c/J \simeq
0.34$. In Table \ref{tab2} we show the estimates of $y_c$ and the
exponent $\gamma$ at various orders. As can be seen, these are quite
consistent at $y_c \simeq 0.473$ and $\gamma \simeq 0.52$. However, this
critical temperature is much higher than the Monte Carlo estimate 0.234
and the corresponding exponent is much lower than the expected Ising
value of 1.75. Therefore we can only conclude that, while the Dlog
Pad{\' e} analysis provides evidence for a physical critical point, the
numerical estimates cannot be taken with any confidence.
We comment further on this in the conclusions.

\begin{figure}
\includegraphics[scale=0.4]{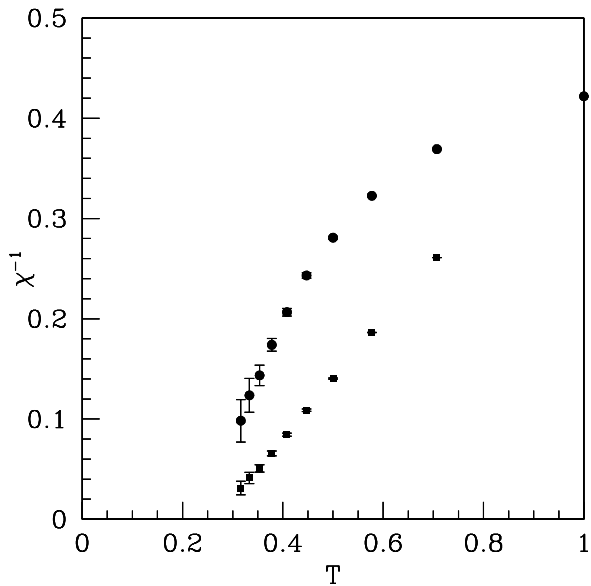}
\caption{
Estimated values of the inverse susceptibility $\beta/\chi$ (circles)
and $\chi^{-1}$ (squares)  
as functions
of temperature $T$ for the 2D quantum compass model (J=1).
}
\label{fig2}
\end{figure}

An alternative approach to analysing our series data is to
evaluate the susceptibility itself at temperatures above
$T_c$, using Pad{\' e} approximants, and to plot the inverse
susceptibility $\chi^{-1}$ versus $T$. 
In Figure \ref{fig2} we plot both $\beta/\chi$, obtained directly from
the series (\ref{eq11}), and $1/\chi$ versus temperature. Both curves
clearly approach zero at $T_c \simeq 0.25$, a value consistent with the
Monte Carlo estimates \cite{wenzel2008,wenzel2010}, and considerably
below our Dlog Pad{\' e} results. It is not possible to obtain accurste
exponent estimates from this procedure, but if we fit our data
points with a simple form $1/\chi = a(T-T_c)^\gamma$ together with the Monte Carlo critical point $T_c = 0.234$ we
obtain $\gamma \simeq 1.3 $, which is at least a good deal closer to the
expected Ising value.

Thus we conclude that the series approach does confirm the
existence of a finite temperature critical point in the
isotropic 2D model, and corroborates the presumably more
accurate Monte Carlo results.

\section{The Simple Cubic Lattice}
\label{sec3}

We now turn to the 3-dimensional model, where no previous results exist.
We use the same approach as for the 2D case, and compute series for the
same quantities. The leading terms of the series for $ln Z$ are
\begin{widetext}
\begin{eqnarray}
 \frac{1}{N} ln Z & = & ln 2 +\frac{1}{2}(x^2 + y^2 = z^2)\beta^2 -\frac{1}{12} (x^4
 + y^4 + z^4 + 8(x^2y^2 + x^2z^2 + y^2z^2)) \beta^4 \nonumber \\
 & & +\frac{1}{45}(x^6 + y^6 + z^6 +30(x^4y^2 +x^2y^4 +x^4z^2 + x^2z^4 + y^4z^2
+ y^2z^4) + 120x^2y^2z^2) \beta^6 \nonumber \\
 & & -\frac{1}{2520}(17(x^8 + y^8 + x^8) +1376(x^6y^2 +x^6z^2 + y^6z^2 + y^6x^2
+ z^6x^2 + z^6y^2) +4344(x^4y^4 + x^4z^4 + y^4z^4) \nonumber \\
 & & \ \ +
14176(x^4y^2z^2+y^4x^2z^2+z^4x^2y^2))
 \beta^8 \nonumber \\
 & & +\frac{1}{14175}(31(x^{10}+y^{10}+z^{10}) +5570(x^8y^2+x^8z^2+y^8x^2+y^8z^2+z^8x^2+z^8y^2))
\nonumber \\
 & & \ \ +40500(x^6y^4 +y^6x^4+x^6z^4+z^6x^4+y^6z^4+z^6y^4) 
+ 120320(x^6y^2z^2+y^6x^2z^2+z^6x^2y^2) \nonumber \\
 & & \ \ +297200(x^4y^4z^2+x^4y^2z^4+x^2y^4z^4)) \beta^{10}
 + \cdots
\label{eq12}
\end{eqnarray}
\end{widetext}
where $x \equiv J_x, y \equiv J_y, z \equiv J_z$.

The susceptibility corresponding to the order parameter
$D_{3d}$ (equation (\ref{eq7})) can be obtained by the substitution $J_x
\rightarrow J_x(1-\lambda), J_y \rightarrow J_y(1-\lambda), J_z \rightarrow
J_z(1+2\lambda)$ in (\ref{eq1}). This definition, of course, introduces a
preferred direction $z$. However in the isotropic limit the resulting series is
unaffected by this.

We have evaluated the series numerically, up to order $\beta^{20}$, and the
coefficients  are
shown in Table \ref{tab3}.

\begin{widetext}
\begin{center}
\begin{table}
\caption{Series coefficients for the isotropic 3D Compass Model}
\begin{tabular}{|cccc|}
\hline
 p &  $\frac{1}{N}ln Z$ & $\chi/(\beta J^2)$ & Q \\
\tableline   
 \ 0 \ &  \ 0.693147180560D+00 \ &  \ 0.600000000000D+01 \ &
 \ 0.600000000000D+01 \ \\
 \ 2 \ &  \ 0.150000000000D+01 \ & -\ 0.600000000000D+01 \ &
 \ 0.180000000000D+02 \ \\
 \ 4 \ & \ -0.225000000000D+01 \ &  \ 0.200000000000D+02 \ &
 \ -0.760000000000D+02 \ \\
 \ 6 \ &  \ 0.673333333333D+01 \ & \ -0.810476190476D+02 \ &
 \ 0.377466666667D+03 \ \\
 \ 8 \ & \ -0.253440476190D+02 \ &  \ 0.367149206349D+03 \ &
 \ -0.200491428396D+04 \ \\
 \ 10 \ &  \ 0.107871111111D+03 \ &  \ -0.178751576719D+04 \ &
 \ 0.110779369318D+05 \ \\
 \ 12 \ & \ -0.496475097002D+03 \ &  \ 0.915575874989D+04 \ &
 \ -0.628679773726D+05 \ \\
 \ 14 \ &  \ 0.241283362972D+04 \ & \ -0.486884786086D+05 \ &
\ 0.363814737295D+06 \ \\
 \ 16 \ & \ -0.122062709687D+05 \ & \ 0.266451000791D+06 \ & \ -0.213714381541D+07 \\
 \ 18 \ &  \ 0.636830117143D+05 \ & & \ 0.127041779973D+08 \\
 \ 20 \ & \ -0.340463327677D+06 \ & & \\ 
\hline
\end{tabular}
\label{tab3}
\end{table}
\end{center}
\end{widetext}

As for the 2D case, the series are dominated by singularities on the negative
$\beta^2$ axis. However, in contrast to the 2D case, Euler transformations
yield no indication of any singularity for real positive $\beta^2$, and thus no
indication of a physical critical point.

\begin{figure}
\includegraphics[scale=0.4]{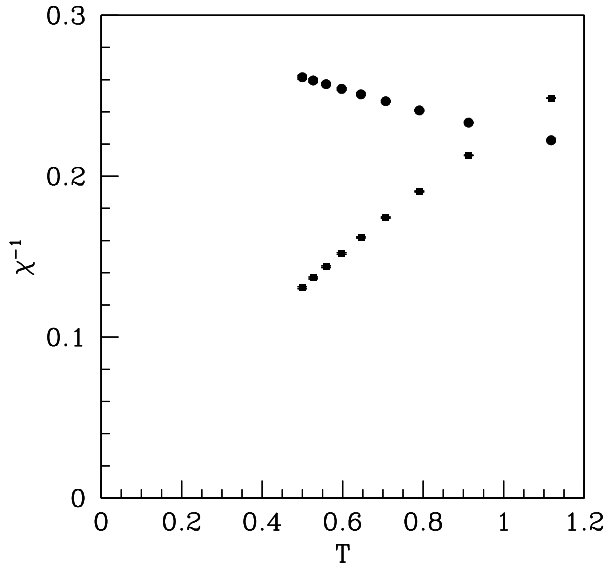}
\caption{
Estimated values of the inverse susceptibility $\beta/\chi$ (circles)
and $\chi^{-1}$ (squares) 
as functions
of temperature $T$ for the 3D quantum compass model (J=1).
}
\label{fig3}
\end{figure}

To test this further we have employed the same strategy as in the previous
section, by evaluating $\chi$ itself at high temperatures, where Pad{\' e}
approximants to the series are well converged, and plotting $\chi^{-1}$ versus
$T$. The results are shown in Figure \ref{fig3}.

We note that the $\beta/\chi$ points are monotonically
increasing, unlike the results for the 2D case - Figure
\ref{fig2}. This indicates that $\chi$ is increasing less rapidly
than $1/T$. The $1/\chi$ values do not indicate a transition at
any finite $T$ but, within the numerical uncertainties, are
consistent with a transition at $T = 0$.

\section{Discussion}
\label{sec4}

The question of the existence of a thermodynamic phase transition in the
quantum compass model on various lattices is of fundamental importance.

The present work is, to our knowledge, the first attempt to address this
problem using the technique of high-temperature series expansions, a standard
method in other contexts.

The series indicate that the analytic structure of thermodynamic functions for
these models is dominated by singularities on the imaginary $\beta$ axis
($\beta = 1/kT$). This is perhaps a reflection of the peculiar
`1-dimensional' nature of the couplings in the model.

Our results for the square lattice are consistent with, albeit less precise
than, recent Monte Carlo results \cite{wenzel2010}. This demonstrates that the high-T
series method does in fact work. However for the cubic lattice we find no
signature of a critical point at finite $T$, and conjecture that there is no
such critical point. At first glance this appears surprising, since the normal
expectation is that the ordered phase will be more robust, and hence $T_c$ will
increase, with increasing dimension. In the case of a simple antiferromagnet, for instance, 
the bond interactions in different directions can be satisfied simultaneously, and reinforce 
each other, so that the tendency to order increases with higher dimension. In the present case, 
however, the bond interactions in different directions pull different ways, and compete 
with each other, so that the tendency to order decreases with higher dimensions.
In one dimension, the 'nematic' order parameter
is non-zero at all finite temperatures; in two dimensions $D_{2d}$ is only non-zero at low 
temperatures; and in three dimensions it appears that $D_{3d}$ is actually zero at all finite
temperatures. It has also been pointed out \cite{khomskii2003,mishra2004}
that in this model thermal fluctuations in fact become larger with increasing
dimension.

The series have proved difficult to analyze, because of the complex
singularities, and gave rather poor estimates of the critical parameters in two
dimensions. A closer investigation of the nature of these singularities may lead to
more precise estimates of the critical parameters; or else higher-order series
coefficients might be necessary.
It is worth noting that the model has also proved difficult to analyze using finite-
size scaling and Monte Carlo methods. An early Monte Carlo calculation \cite{tanaka2007} on 
lattices of up to 20 x 20 sites with periodic boundary conditions also gave a critical point 
about 36\% too high. Wenzel {\it et al.} \cite{wenzel2010} showed that the use of special `screw periodic'
boundary conditions on lattices up to 42 x 42 was required to produce the estimates quoted 
earlier.

\begin{acknowledgments}
We are grateful for the
computing resources provided by the Australian Partnership for
Advanced Computing (APAC) National Facility.
\end{acknowledgments}

\end{document}